\documentclass[letterpaper, 10 pt, journal, twoside]{IEEEtran}  % Comment this line out if you need a4paper

\IEEEoverridecommandlockouts                              % This command is only needed if 
\usepackage{graphics} % for pdf, bitmapped graphics files
\usepackage{epsfig} % for postscript graphics files
\usepackage{amsmath} % assumes amsmath package installed
\usepackage{amssymb}  % assumes amsmath package installed
\usepackage{mathtools}
\usepackage{bm}
\newcommand{\vect}[1]{\boldsymbol{\mathbf{#1}}}
\usepackage{units}
\usepackage{siunitx}
\usepackage{cite}
\usepackage{balance}
\usepackage{color}
\usepackage{url}

\usepackage[printonlyused]{acronym}

\acrodef{ASL}{Autonomous Systems Lab}
\acrodef{UAV}{Unmanned Aerial Vehicle}
\acrodef{VTOL}{Vertical Take-Off/Landing}
\acrodef{CG}{center of gravity}
\acrodef{DoF}{degrees of freedom}
\acrodef{IMU}{Inertial Measurement Unit}
\acrodef{IG}{initial guess}
\acrodef{TM}{trim-map}
\acrodef{PID}{Proportional-Integral-Derivative}
\acrodef{RHS}{right-hand side}
\acrodef{CAD}{computer-aided design}
\acrodef{TWV}{tiltwing vehicle}
\acrodef{ACS}{attitude control system}
\acrodef{CCS}{cruise control system}
\acrodef{DI}{dynamic inversion}
\acrodef{RC}{radio controller}

\usepackage{pgfplots} 
\usetikzlibrary{external}
\usepackage{tikz}
\usepackage[final]{pdfpages}

\newcommand{\added}[1]{\textcolor{black}{#1}}

\newcommand{\deleted}[1]{}

\newcommand{\short}[1]{\textcolor{black}{#1}}

\title{Attitude- and Cruise Control of a VTOL Tiltwing UAV} %Use only for final RAL version.

\author{David Rohr, Thomas Stastny, Sebastian Verling and Roland Siegwart% <-this % stops a space
\thanks{\added{Manuscript received: 02, 24, 2019; Revised -; Accepted 04, 14, 2019.}}
\thanks{\added{This paper was recommended for publication by Editor Jonathan Roberts upon evaluation of the Associate Editor and Reviewers' comments. The authors would like to thank Dufour Aerospace (https://dufour.aero) for initiating and supporting this project.}}
\thanks{The authors are with the Autonomous Systems Lab, ETH Zurich, Leonhardstrasse 21, 8092 Zurich, Switzerland, {\tt\small \added{<firstname.lastname>@mavt.ethz.ch}}}%
\thanks{\added{Digital Object Identifier (DOI): see top of this page.}} }

\begin{document}
\null%
\includepdf[pages=-]{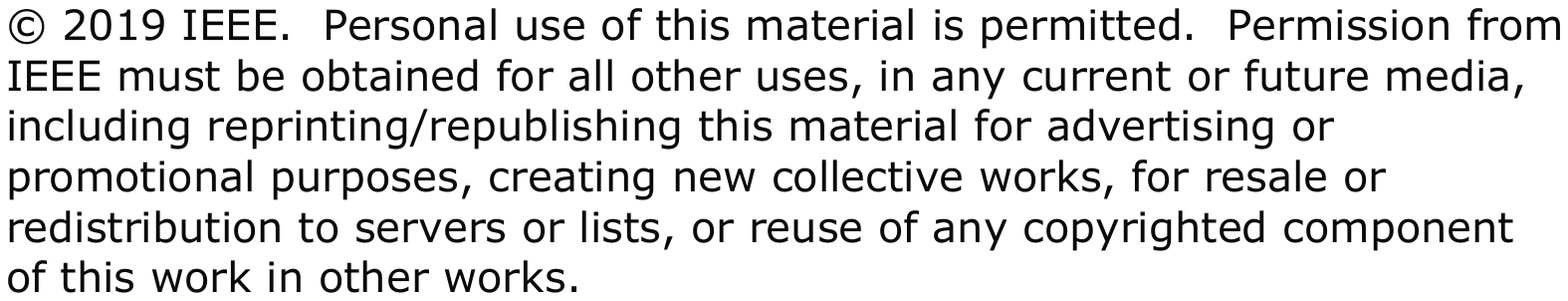}
\setcounter{page}{1}

\maketitle
%\thispagestyle{empty} % Comment or remove this line for final RAL version.
%\pagestyle{empty}     % Comment or remove this line for final RAL version.

%%%%%%%%%%%%%%%%%%%%%%%%%%%%%%%%%%%%%%%%%%%%%%%%%%%%%%%%%%%%%%%%%%%%%%%%%%%%%%%%

\begin{abstract}
This paper presents the mathematical modeling, controller design, and flight-testing of an over-actuated Vertical Take-off and Landing (VTOL) tiltwing \ac{UAV}. 
Based on simplified aerodynamics and first-principles, a dynamical model of the UAV is developed which captures key aerodynamic effects including propeller slipstream on the wing and post-stall characteristics of the airfoils.
The model-based steady-state flight envelope and the corresponding trim-actuation is analyzed and the overactuation of the UAV solved by optimizing for, e.g., power-optimal trims. 
The developed control system is composed of two controllers: First, a low-level attitude controller based on dynamic inversion and a daisy-chaining approach to handle allocation of redundant actuators. Secondly, a higher-level cruise controller to track a desired vertical velocity. 
It is based on a linearization of the system and look-up tables to determine the strong and nonlinear variation of the trims throughout the flight-envelope. 
We demonstrate the performance of the control-system for all flight phases (hover, transition, cruise) in extensive flight-tests.
\end{abstract}


%%%%%%%%%%%%%%%%%%%%%%%%%%%%%%%%%%%%%%%%%%%%%%%%%%%%%%%%%%%%%%%%%%%%%%%%%%%%%%%%

\section{INTRODUCTION}
\begin{figure}[tb]
	\centering
	\includegraphics[width = 1\linewidth]{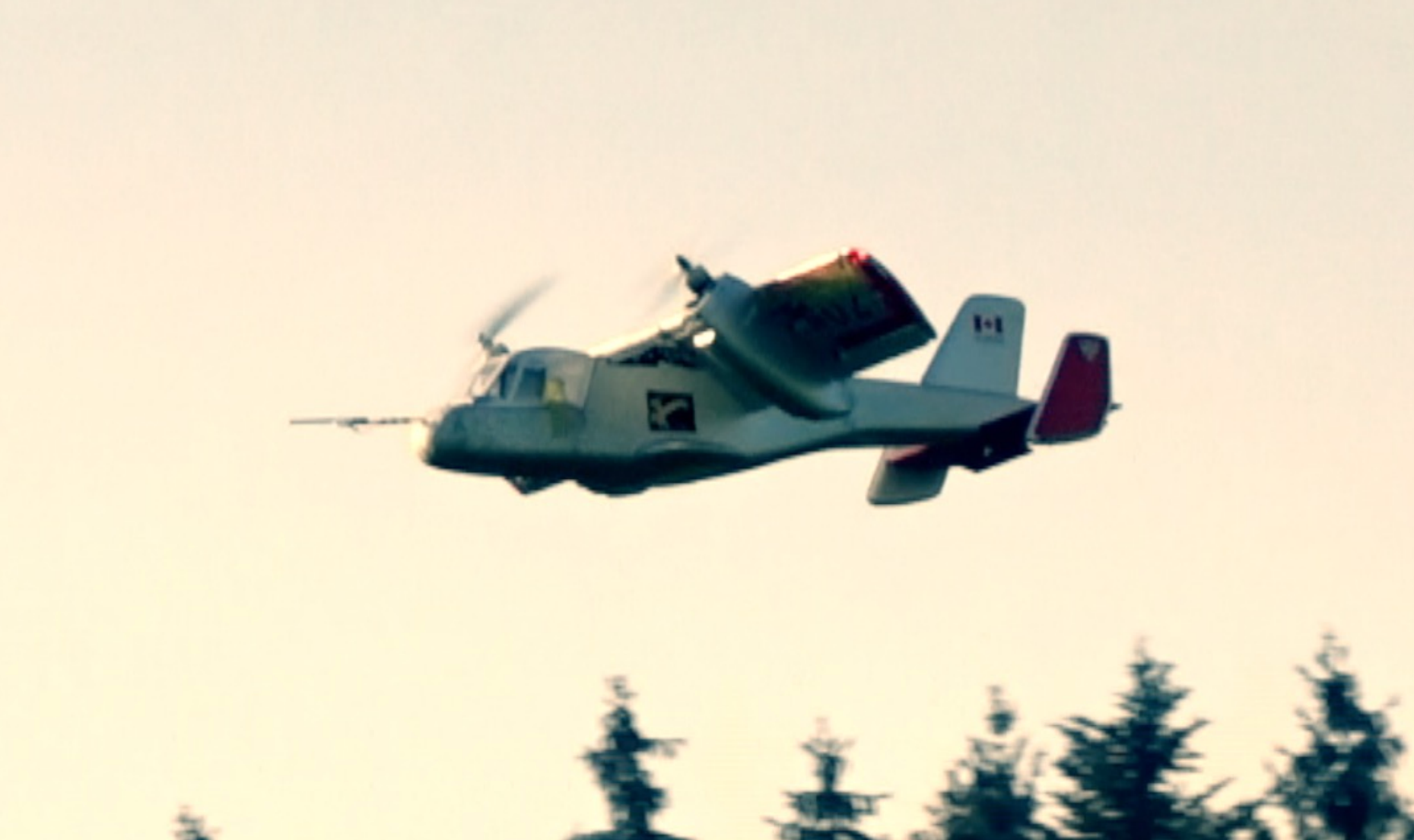}
	\caption{The tiltwing UAV on which the presented control system is implemented, here shown in transition mode with partially tilted wing.}
	\vspace{-0.5cm}
	\label{fig:cl84_airborne}
\end{figure}

\IEEEPARstart{U}{nmanned} aerial vehicles (UAVs) are extensively investigated in the robotics community. Over the last decades, various designs evolved to meet the requirements of specific mission profiles. Fixed-wing aircraft designs offer high endurance, large range, and high speeds while rotary-wing platforms such as the popular \emph{multirotor} feature high maneuverability, hover- and \ac{VTOL} capabilities.  
There is increasing interest in the development of highly versatile, so-called ``hybrid'' \acp{UAV} that can operate both as fixed- and rotary-wings and thus combine the benefits of the respective designs \cite{Saeed2015}. Examples are the \emph{tiltwing} presented in this work, the \emph{tiltrotor} and the \emph{tailsitter}. A \ac{TWV} features a wing that rotates together with the propulsion system between a horizontal (cruise-mode) and upright- (hover-mode) position (Fig. \ref{fig:cl84_schematic}). 
At intermediate wing-tilt angles, the lift-force resolves into contributions of both the propulsion system and the airfoils, leading to a blend of fixed- and rotary wing operations (transition mode), see Fig. \ref{fig:cl84_airborne}.
Compared to the tailsitter design \cite{Verling2016,Ritz2017}, \acp{TWV} benefit from the fuselage remaining horizontal. Versus the tiltrotor, they feature improved stall characteristic and more effective wing-born lift due to the continuous immersion of the wing in the well-aligned propeller slipstream.
 
The large flight-envelope of a \ac{TWV} imposes challenging requirements on the flight-control system. The transition phase is characterized by strong nonlinearities in the dynam- ics of the aircraft. These result from the interaction of the wing and the propeller slipstream, the wing’s wide range in angles of attack, and, generally, the large variation of trim-settings throughout the flight envelope. For the \ac{TWV} presented in this work, an additional control challenge is given by actuator redundancy that leads to an overactuation for both attitude- and cruise control. 

\subsubsection{Related Work}
Existing work on design, modeling and control of tiltwing UAV's considers tandem-wing \cite{Uchiyama2019,Mcswain2017,Muraoka2013} and single-wing vehicles \cite{Nikolakopoulos2016,Dickeson2007,Hartmann2016}.  
Employed control systems are either unified \cite{Dickeson2007,Hartmann2016} or switch between different controllers for hover, transition, and cruise \cite{Ostermann2012}. 
For both attitude- and velocity control, decoupled PID and full state feedback LQR-architectures are reported \cite{Mcswain2017,Hartmann2017_2,Hartmann2016}. They are typically combined with local linearizations and gain-scheduling to address the strong non-linearities. 
Examples of $H_{\infty}$-based attitude- and cruise control are found as well \cite{Dickeson2005,Uchiyama2019}. 
A popular non-linear control technique involves \ac{DI}\cite{Khalil2002} and is used both for tailsitters \cite{Ritz2017,Verling2016} and tandem \acp{TWV} \cite{Uchiyama2019}. It enables reference-model following but requires an accurate model to estimate state-dependent moments and forces.  
High-fidelity models are required to address the complex transition phase and typically consider the prominent propeller slipstream interaction with the wing \cite{Hartmann2017_2}. Instead of modeling, \cite{Hartmann2016} describes a control system that is based exclusively on state- and control derivatives obtained from wind-tunnel testing, \cite{Verling2016} and \cite{Ritz2017} introduce lumped-parameter models to fit experimental data for a flying-wing tailsitter. 

\subsubsection{Contribution}
In this work, we present a global, model-based control system that tracks the full desired attitude and vertical airspeed in all flight phases. The \ac{ACS} is based on i) a high-fidelity model built from first principles, ii) dynamic inversion and iii) a daisy-chaining approach to handle overactuation. This combination is novel in its application to single-wing \ac{TWV}s. 
The good performance of the \ac{ACS} in the flight-tests indicates that the proposed model structure reasonably trades-off between i) \ac{DI}-required fidelity and ii) low-computational complexity to remain tangible for use on micro-controllers.    
The developed \ac{CCS} employs a linearized approach similar to that outlined in \cite{Hartmann2016}, i.e., it relies on look-up \acp{TM} to determine the nonlinear trim-actuation. However, contrary to \cite{Hartmann2016} where wind-tunnel based \acp{TM} are used, we rely on model-based \acp{TM} obtained by offline full-state optimization to systematically handle non-uniqueness of the trims. Additionally, the \ac{CCS} presented includes feedback control \deleted{to increase robustness against modeling errors and disturbances}\added{to account for modeling errors and to attenuate disturbances}. The resulting vertical velocity tracking accuracy and -range improves on the data presented in \cite{Hartmann2016}, thus rendering modeling with velocity feedback a valuable alternative to laborious wind-tunnel testing.

\subsubsection{Outline}
The remainder of the paper is structured as follows: In Section~\ref{sec: system_description}, the system is introduced, followed by the modeling in Section~\ref{sec: system_modeling}. Optimal trim-actuation is analyzed in Section~\ref{sec: trim_analysis} and forms the basis for the \ac{CCS}. Section~\ref{sec: attitude_control} and~\ref{sec: cruise_control} introduce the attitude- and cruise-control architectures, respectively. The performance of the controllers is demonstrated in flight experiments presented in Section~\ref{sec: experimental_verification}. Finally, an outlook is given in Section~\ref{sec: conclusion_future_work}. 

\section{SYSTEM DESCRIPTION}\label{sec: system_description}
The employed tiltwing-UAV is a commercially available radio-controlled (RC) aircraft \cite{RC_CL84}. It is a replica of the Canadair \mbox{CL-84} manned tiltwing aircraft which flew in the 1960's and comes fully equipped with all required actuators including a tilt-mechanism for the wing. It features a wing-span of \SI{0.94}{m} and its take-off mass amounts to \SI{1.9}{kg}. In cruise configuration, flights of up to  $\sim\SI{20}{min}$ are possible while in hover, endurance is limited to $\sim\SI{5}{min}$ (battery: \SI{14.8}{V}, \SI{3800}{mAh}).

\subsection{Avionics}\label{ssec: avionics}
In order to implement our own flight-control system, the UAV is refitted with the Pixhawk Autopilot \cite{pixhawk} running the PX4 autopilot software \cite{Meier2015}. The Pixhawk provides a six-axis Inertial Measurement Unit (IMU) (3-axis accelerometer + 3-axis gyroscope), a 3-axis magnetometer, and a barometer. Additionally, the system is complemented with a differential-pressure airspeed-sensor and a GNSS-module. All sensor data is fused in the ready-to-use state-estimation available within PX4, resulting in attitude, altitude and airspeed estimates that are subsequently used in the feed-back flight-control system.  
 
\subsection{Actuation Principle}\label{ssec: actuation_principle} 
\begin{figure}[tb]
	%\vspace{-0.5cm}
	\centering
	\includegraphics[width = 1\linewidth]{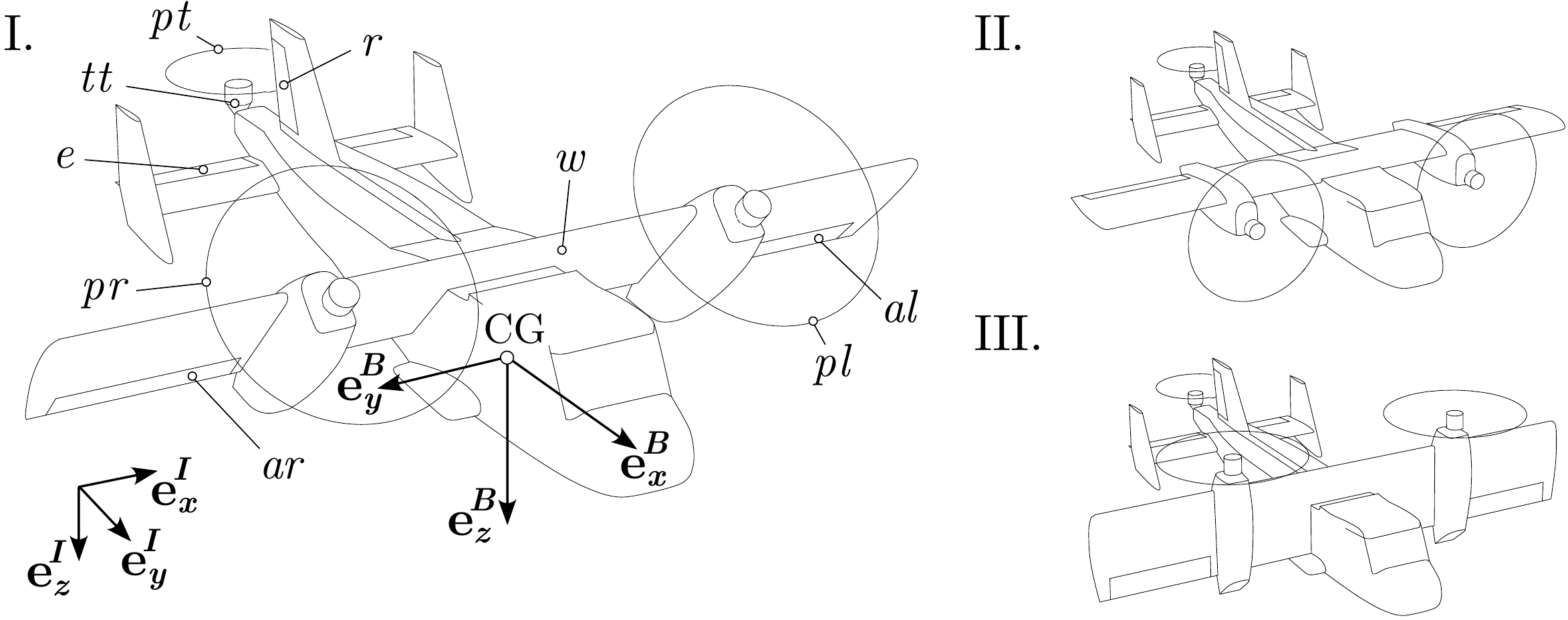}
	\caption{Schematic of the tiltwing UAV in the different flight-phases hover (III), transition (I) and cruise (II). The body-fixed frame of reference is located in the center of gravity (CG) and denoted with the superscript ${B}$, the inertial, earth-fixed frame is labeled with the superscript ${I}$.}
	\vspace{-0.5cm}
	\label{fig:cl84_schematic}
\end{figure}

Fig.~\ref{fig:cl84_schematic} depicts the different available actuators. Their function depends on the flight phase and the configuration of the UAV: 

In \emph{hover flight}, roll and pitch are controlled by thrust-vectoring of the main-propellers (\emph{pl, pr}) and the tail-propeller (\emph{pt}). Yaw is actuated by tilting the tail-propeller thrust vector around the body x-axis (\emph{tt}). 
Redundantly, yaw-moment can also be generated by differential deflection of the slipstream-immersed ailerons (\emph{al, ar}). Horizontal maneuvering is performed by tilting the UAV, and hence the net thrust-vector, into the desired direction. Climbing and sinking is achieved by collective throttling of all propellers.

In \emph{cruise flight}, standard fixed-wing controls apply, i.e., roll, pitch and yaw are controlled with the ailerons, elevator (\emph{e}) and rudder (\emph{r}), respectively. Additionally, yaw- and negative pitch moment can be generated by differential throttle on the main-propellers and the tail-propeller thrust, respectively. Again, this provides redundancy and illustrates the overactuation for attitude control. Airspeed and climb-rate are controlled by coordinating main-propeller thrust and pitch-angle of the UAV.

In the \emph{transition} phase, the control strategies overlap, e.g., rolling and yawing both require simultaneous thrust vectoring and aileron deflection. Horizontal- and vertical velocity control includes combined wing-tilt actuation, pitch-angle- and throttle selection. This combination is not necessarily unique, hence, overactuation is again present:   
For example, hovering is possible with every wing-tilt angle ($\zeta_w$) and fuselage-pitch ($\theta$) combination that leads to the main-thrust vector pointing upward, i.e., $\zeta_w+\theta\approx\ang{90}$.

\subsection{Nomenclature}\label{ssec: nomenclature}
\short{
\added{For system analysis, we introduce a body-fixed forward-right-down frame of reference located in the UAV's \ac{CG} (cf. Fig.~\ref{fig:cl84_schematic}). \ac{CG} variation upon tilting of the wing is neglected, it amounts to $\pm$ \SI{1}{cm} along the body z-axis w.r.t. the \ac{CG} used for modeling.}
Vector-quantities are written in bold-face and denoted with lower pre-script $\mathcal{B}$ and $\mathcal{I}$ when expressed in body- and inertial-frame, respectively (cf. Fig.~\ref{fig:cl84_schematic}). 
Normalized actuator control inputs are denoted with $\delta_i$, actual actuator positions with $\zeta_i$, and propeller speeds with $\eta_i$. The subscript defines the actuator and will be introduced when used.
} 

\section{SYSTEM MODELING}\label{sec: system_modeling}
The UAV is modeled as a single rigid body whose translational and angular dynamics are driven by the net force and moment acting on it. With the Newton-Euler equations and rigid-body kinematics, the position ($\mathbf{x}$), translational velocity ($\mathbf{v}$), attitude ($\mathbf{R}_\mathcal{IB}$) and angular rate ($\bm{\omega}$) of the UAV are defined by
\begin{equation}\label{eq: EoM}
\begin{array}{rcl}
\prescript{}{\mathcal{I}}{\dot{\vect{x}}}&=&\prescript{}{\mathcal{I}}{\vect{v}}\\[3pt]
m\prescript{}{\mathcal{I}}{\dot{\vect{v}}}&=&m\prescript{}{\mathcal{I}}{\vect{g}} + \mathbf{R}_{\mathcal{IB}}\cdot\prescript{}{\mathcal{B}}{\vect{F}}\\[3pt]
\dot{\mathbf{R}}_{\mathcal{IB}}&=& \mathbf{R}_{\mathcal{IB}}\left[\prescript{}{\mathcal{B}}{\bm{\omega}}\right]_\times\\[3pt]
\prescript{}{\mathcal{B}}{\vect{I}}
\prescript{}{\mathcal{B}}{\dot{\bm{\omega}}}&=&\prescript{}{\mathcal{B}}{\mathbf{M}} - \prescript{}{\mathcal{B}}{\bm{\omega}}\times
\prescript{}{\mathcal{B}}{\mathbf{I}}
\prescript{}{\mathcal{B}}{\bm{\omega}}\\[3pt]
\end{array}
\end{equation}
where $m$ denotes the UAV's total mass and $\prescript{}{\mathcal{B}}{\mathbf{I}}$ the UAV's moment of inertia expressed in $\mathcal{B}$. The gravitational acceleration is given by $\mathbf{g}$, $\left[\prescript{}{\mathcal{B}}{\bm{\omega}}\right]_\times$ denotes the skew-symmetric cross-product matrix of $\prescript{}{\mathcal{B}}{\bm{\omega}}$. We express the attitude ($\mathbf{R}_\mathcal{IB}$) with a rotation-matrix mapping from body- to inertial frame. The net aerodynamic force $\mathbf{F}$ and moment $\mathbf{M}$ are formed by accumulating the contributions of the different components, i.e., wing, stabilizers, fuselage and propellers.
Tiltwing-specific aerodynamic effects to be modeled include the forces and moments generated by \short{i) the airfoils subject to full $\pm$180$^\circ$ free-stream angle of attack, ii) the propeller-slipstream effects on airfoils located downstream of the propellers, and iii) the propellers facing different inflow-conditions throughout the flight envelope.}

\subsection{Propellers}\label{ssec: propeller_model} 
With $\vect{v_a}$ the air-relative velocity (airspeed) of the UAV's CG, the local airspeed $\vect{u_a}$ at the propeller hub is given by  
\begin{equation}
\vect{u_a}(\vect{r_{p}}) = \vect{v_a} + \bm{\omega}\times\vect{r_{p}} = V_{\parallel,\infty}\vect{p}_{\parallel} + V_{\perp,\infty}\vect{p}_{\perp} 
\end{equation}
with $\vect{r_{p}}$ the CG-relative position of the propeller hub. The local airspeed resolves in an axial ($V_{\parallel,\infty}$) and radial \mbox{($V_{\perp,\infty}\geq0$)} free-flow component, $\vect{p}_{\parallel}$ and $\vect{p}_{\perp}$ are unit vectors pointing in propeller forward and radial direction, respectively. According to \cite{Selig2010_2,Martin2010}, the net force of propeller $p$ is composed of the thrust $T$ and normal force $N$:
\begin{equation}\label{eq:Thrust}
\prescript{}{\mathcal{B}}{\vect{F}_p}  = \underbrace{\rho \eta^2 D^4 C_T(J)}_{T}\prescript{}{\mathcal{B}}{\vect{p}_\parallel} - \underbrace{\eta\mu_N V_{\perp,\infty}}_{N}\prescript{}{\mathcal{B}}{\vect{p}_\perp}
\end{equation}
with the thrust-constant $C_T$ depending on the propeller advance-ratio $J$ and a lumped-parameter constant $\mu_N>0$. Further, $\rho$, $\eta$, and $D$ denote air density, propeller speed and propeller diameter, respectively.   

The reactive propeller-moment due to the air-drag of the propeller blades amounts to
\begin{equation}\label{eq: drag torque}
\prescript{}{\mathcal{B}}{\vect{M}_p} = -\rho \eta^2 D^5 C_Q(J)\varepsilon\prescript{}{\mathcal{B}}{\vect{p}_\parallel}
\end{equation}
with $C_Q$ the torque-constant. The propeller turning direction is determined by $\varepsilon\in\left\lbrace -1,1 \right\rbrace$ where $\varepsilon=1$ if the turning direction is positive along $\vect{p}_\parallel$ (right-handedness) and vice-versa. $C_T$ and $C_Q$ are approximated as affine function of $J$ (cf. \cite{Selig2010_2}). For simplicity, other components, such as the propeller rolling moment, are neglected.

\subsection{Airfoils}\label{ssec: airfoil_model}
\begin{figure}[b]
	\vspace{-0.5cm}
	\centering
	\includegraphics[width=0.45\linewidth]{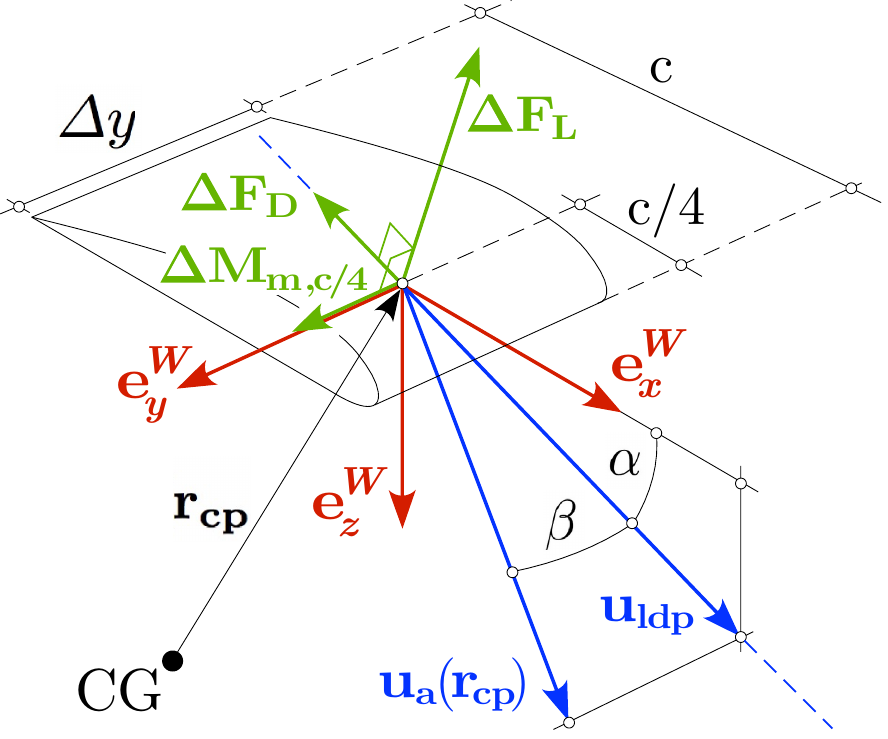}
	\caption{Schematic of an airfoil segment with a local frame of reference 
		\mbox{$\mathcal{W} = (\vect{e_x^W},\vect{e_y^W},\vect{e_z^W})$} and the quantities involved to compute the aerodynamic forces and -moments.}
	%\vspace{-0.5cm}
	\label{fig: wing def}
\end{figure}

We divide the wing and stabilizers into multiple span-wise segments to account for the different inflow conditions which depend on $\bm{\omega}$ and, for wing segments located behind a propeller, the propeller slipstream velocity $\vect{w}$. Resulting forces and moments are calculated at the center of pressure $\vect{r_{cp}}$ of each segment, see Fig.~\ref{fig: wing def}. The local airspeed $\vect{u_a}$ is given by
\begin{equation}
\vect{u_a}(\vect{r_{cp}}) = \vect{v_a} + \bm{\omega}\times\vect{r_{cp}}\quad(+ \vect{w}) 
\end{equation}
For simplicity, the spatial evolution of the propeller wake \cite{Selig2010_2,Khan2014} is neglected and the induced velocity $\vect{w}$ approximated by the value at the corresponding propeller-hub. From disk actuator theory \cite{Selig2010_2}:
\begin{equation}\label{eq:indVel}
\vect{w} = \vect{p}_\parallel\frac{1}{2}\left[-V_{\parallel,\infty} + \sqrt{V_{\parallel,\infty}^2 + \left(\frac{2T}{\rho A}\right)}\right]
\end{equation}
with \mbox{$A=\pi D^2/4$} the propeller-disk area. Segment-wise lift ($\vect{\Delta F_L}$), drag ($\vect{\Delta F_D}$) and moment ($\vect{\Delta M}$) contributions are finally obtained by
\begin{equation}\label{eq: airfoil_aerodynamic}
\begin{array}{rcl}
\vect{\Delta F_L}&=&C_L(\alpha,\zeta_{cs})\cdot\frac{1}{2}\rho V^2\cdot c\cdot \Delta y \cdot \vect{e_L}\\[3pt]
\vect{\Delta F_D}&=&C_D(\alpha,\zeta_{cs})\cdot\frac{1}{2}\rho V^2\cdot c\cdot \Delta y \cdot \vect{e_D}\\[3pt]
\vect{\Delta M_{m,c/4}}&=&C_M(\alpha,\zeta_{cs})\cdot\frac{1}{2}\rho V^2\cdot c^2\cdot \Delta y \cdot \vect{e_y^W}\\[3pt]
\end{array}
\end{equation}
where the definition of most quantities is illustrated in Fig.~\ref{fig: wing def}. Lift- and drag direction are denoted by $\vect{e_L}$ and $\vect{e_D}$, respectively, and $V=\|\vect{u_{ldp}}\|$ (cf. Fig.~\ref{fig: wing def}). The aerodynamic coefficients $C_L$, $C_D$, $C_M$ depend on the angle of attack $\alpha\in[-\pi,\pi]$ and, if control surfaces (CS) are present, the CS deflection $\zeta_{cs}$. Simple linear and quadratic relations are employed if the segment is not stalled \mbox{($\alpha_{s^-} < \alpha < \alpha_{s^+}$)}. In post stall ($\alpha_{s^-} \gg \alpha, \alpha \gg \alpha_{s^+}$), the wing is assumed to behave like a flat plate (\emph{fp}) and we approximate the coefficients by:
\begin{equation}\label{eq: flatplate_aeroForce}
\begin{array}{rcl}
C_L^{fp}&=&C_{L,\pi/4}^{fp}\cdot sin(2\alpha)\\[3pt]
C_D^{fp}&=&C_{D,min}^{fp}+(C_{D,\pi/2}^{fp}-C_{D,min}^{fp})\cdot \text{sin}(\alpha)^2\\[3pt]
C_M^{fp}&=&-C_{M,max}^{fp}\cdot \text{sin}(\text{sgn}(\alpha)\cdot\alpha^2/\pi)\\[3pt]
\end{array}
\end{equation}
to match reported experimental data, e.g., \cite{Ortiz2012, Hoerner1965}. %Post-stall, we neglect effects of control-surfaces.
Close to the stall angles ($\alpha_{s^-}$, $\alpha_{s^+}$), \short{interpolation between both models yields a smooth changeover}. %The resulting lift- and drag curves are shown in Fig. XX.

\subsection{Fuselage}
Modeling of the fuselage follows a simplified approach known as ``quadratic aerodynamic form'' (cf. \cite{Heffley1988}, p.20): 
\begin{equation}
	\begin{array}{c}
	\prescript{}{\mathcal{B}}{\vect{F_f}}=-\frac{\rho}{2}\left(\vect{e}_x^B C_{D,x}^f V_{uu}+\vect{e}_y^B C_{D,y}^f V_{vv}+\vect{e}_z^B C_{D,z}^f V_{ww}\right)\\[3pt]
	\prescript{}{\mathcal{B}}{\vect{M_f}}=0\\[3pt]
	V_{uu} = u|u|,\quad V_{vv} = v|v|, \quad V_{ww} = w|w|
	\end{array}
\end{equation}
with $(u,v,w)^T = \prescript{}{\mathcal{B}}{\vect{v_a}}$ and $C_{D,i}^f$ the drag coefficients of the fuselage subject to $i$-axis-aligned airflow.

\subsection{Parameter Identification}
The introduced parameters are either identified (experimental assessment of static thrust- and moment curves of the propulsion systems and CAD-based approximation of inertial properties) or estimated based on values obtained from literature (all airfoil-related data). 

\section{TRIM ANALYSIS}\label{sec: trim_analysis}
Cruise-control system development is preceded by assessing the steady-state flight envelope, i.e., the set of operating points for which a dynamic equilibrium exists. 
We restrict the formulation of the cruise-controller and the system analysis to the 2-dimensional, longitudinal dynamics. An operating point can therefore be defined by the flight-path angle $\gamma$ and the airspeed magnitude $v_a$. 
The point is declared steady feasible if for the given pair $(v_a,\gamma)$ there exists a trim-pitch $\theta^t$ and a set of trim-actuations $\vect{u}^t$
\begin{equation}
\theta^t(v_a,\gamma),\quad\vect{u}^t(v_a,\gamma) = \begin{bmatrix}\delta_w^t&\delta_{pl,r}^t&\delta_{al,r}^t&\delta_e^t&\delta_{pt}^t\end{bmatrix}
\end{equation}
such that
\begin{equation}
\bm{\omega} = \dot{\bm{\omega}} = \dot{\vect{v}} = \vect{0}.
\end{equation}
The superscript $t$ relates to the trim setting and the subscripts of the $\delta$'s specify the actuator (cf. Fig.~\ref{fig:cl84_schematic}). 
If no such actuation/pitch-angle exists, the operating point cannot be stabilized by any control system. The mapping $\mathcal{T}: (v_a,\gamma) \rightarrow (\vect{u^t},\theta^t)$ is referred to as trim-map. It is calculated in a discrete form to serve as look-up table for feed-forward actuation of the  wing-tilt, throttle, and pitch for cruise control (cf. Section~\ref{ssec: trim_map_ff}). With $\mathcal{T}$ not necessarily unique (overactuation, see Section~\ref{ssec: actuation_principle}), selection from a set of feasible trims follows an optimization which regards the use of the trims for feed-forward actuation in cruise control. This differs from the approach in \cite{Hartmann2016}, where $\theta^t$ is imposed to render the trims unique---this potentially constrains the solution space, i.e., the extent of the assessable flight envelope. Furthermore, in \cite{Hartmann2016} $\mathcal{T}$ is entirely based on wind-tunnel data, i.e., no model or optimization is involved.

\subsection{Problem Formulation and Optimization}\label{ssec: trim_optimization}
The trim-map is calculated offline in a nonlinear, constrained optimization which minimizes translational ($\dot{\vect{v}}$) and angular ($\ddot{\theta}$) unsteadiness and, simultaneously, seeks to reduce a user-defined cost-function $q$ which is included to render the trim solution unique:
\begin{equation}
\begin{array}{l}
\underset{\mathbf{u^t},\theta^t}{\text{min}} 
\left(\dot{\vect{v}}^T \vect{Q_v}\dot{\vect{v}} + Q_{\theta}\ddot{\theta}^2 + q(\mathbf{u^t},\theta^t)\right)\quad\text{s.t.}\\[6pt]
%\multicolumn{1}{l}{\text{s.t.}}\\[6pt]

\multicolumn{1}{c}{\prescript{}{\mathcal{I}}{\dot{\vect{v}}}=\prescript{}{\mathcal{I}}{\vect{g}} + \frac{1}{m}\mathbf{R}_{\mathcal{IB}}\cdot\prescript{}{\mathcal{B}}{\vect{F}(\mathbf{u^t},\theta^t)}}\\[6pt]

\multicolumn{1}{c}{\prescript{}{\mathcal{B}}{\dot{\bm{\omega}}}=\prescript{}{\mathcal{B}}{\vect{I}}^{-1}\prescript{}{\mathcal{B}}{\mathbf{M}(\mathbf{u^t},\theta^t)}}\\[6pt]

\multicolumn{1}{c}{-\pi/2\leq\theta^t\leq\pi/2, \quad \vect{u^t}\in\mathbb{U}}%\\[6pt]

\end{array}
\end{equation}
with $\vect{Q_v}$, $Q_\theta$ positive definite weightings and $\mathbb{U}$ the set of admissible, non-saturated actuator inputs. The equality constraints follow from the system dynamics (\ref{eq: EoM}), $\vect{F}$ and $\vect{M}$ further depend on $(v_a,\gamma)$. In $q\geq0$ we include penalties on \short{i) net power-consumption, ii) control-surface saturation, iii) deviation from a desired pitch-angle $\theta^*$ and iv) deviation from solutions of close operating points to penalize discontinuous trim-maps and, thus, prevent discrete switching of the feed-forward trim values in the cruise controller.}

At every $(v_a(i),\gamma(j))$ contained in the trim-map, we performed the optimization using the \emph{lsqnonlin} solver of the Matlab optimization toolbox \cite{MatlabOTB}. Finally, the resulting steadiness ($\dot{\vect{v}},\ddot{\theta}$) was thresholded to decide upon incorporation of $(v_a(i),\gamma(j))$ in the steady flight envelope. It is worth noting that the resulting steady flight envelope is generally a conservative estimate due to the risk of the solver getting trapped in a local optimum or too much weight being put on minimizing the additional cost $q$. To minimize the risk of locally optimal solutions, the solver requires appropriate initial guesses (IG).

\subsection{Initial Guess Generation}\label{ssec: IG_gen}
We devise an iterative procedure to generate IGs during build-up of the trim-map: At every operation point $(v_a(i),\gamma(j))$ in the trim-map, the optimization is solved once with every available solution of the neighboring operation points as IG (eight in total for a Cartesian grid). The steady-feasible solution $(\vect{u^t},\theta^t)$ which yields the lowest cost is adopted as preliminary trim at $(v_a(i),\gamma(j))$. If, in a subsequent iteration, a neighboring point manages to further lower its cost with a new solution, the trim at $(v_a(i),\gamma(j))$ is revisited with this solution as IG and adjusted upon improvement. This procedure is conducted at every point in the map for multiple iterations until the solutions do not change anymore. At this instant, mutually lowest costs are achieved among neighboring points in the map. 

The procedure requires at least one $(v_a(i),\gamma(j))$ to be solved in advance, its IG is provided manually. If the grid points $(v_a(i),\gamma(j))$ are spaced close enough and assuming sufficient smoothness of the optimal trim-map, this procedure provides IGs which are already close to the actual solution. Further, it fosters propagation of good solutions through the map: though not guaranteed, a globally optimal solution at $(i,j)$ might render locally optimal solutions in $(i + m, j + n)$ globally optimal as well.

\subsection{Results}\label{ssec: trim_analysis_results}
\begin{figure}[t]
	\centering
	%\vspace{-0.25cm}
	\includegraphics[width = 1\linewidth]{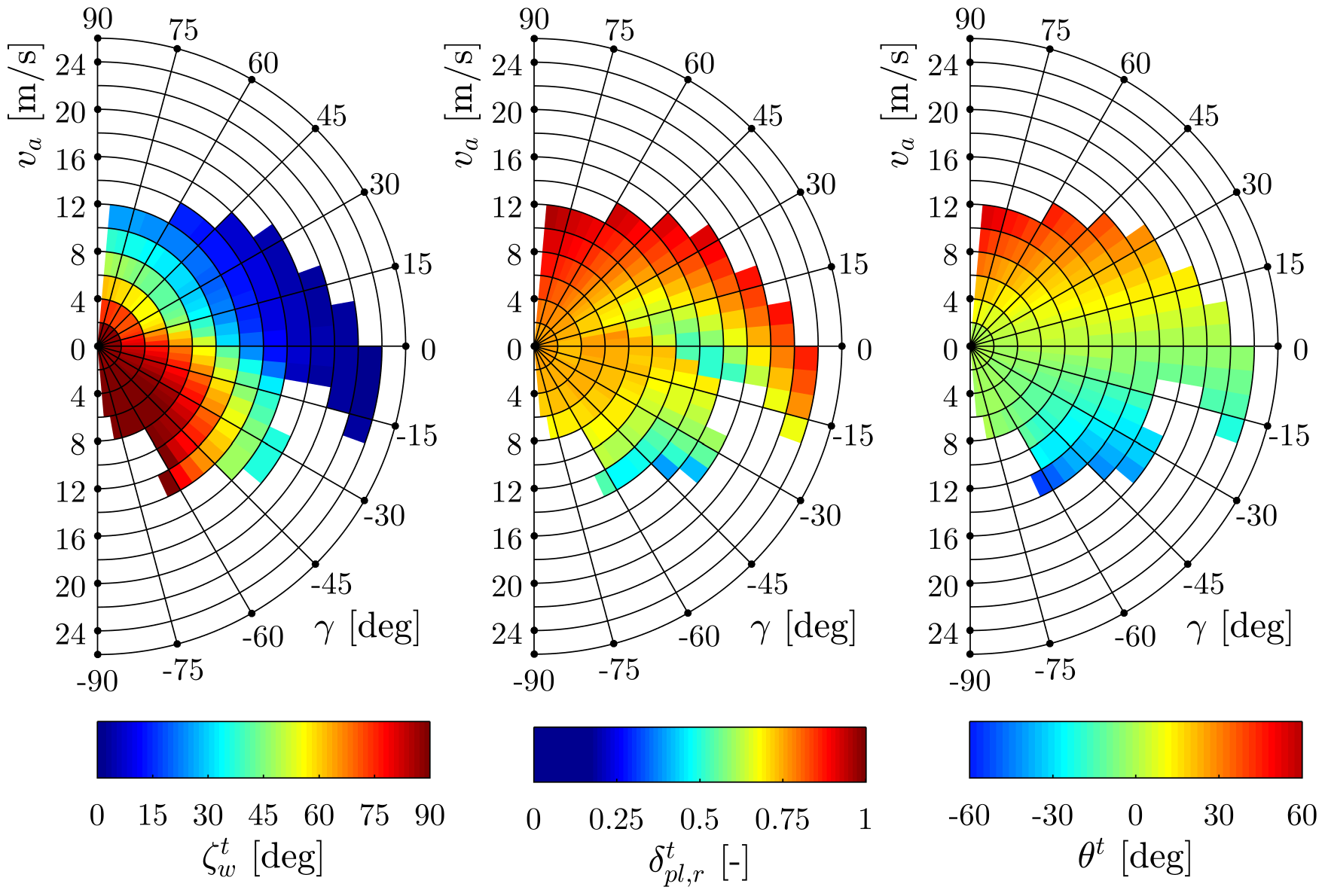}
	\vspace{-0.5cm}
	\caption{Trim-maps obtained from the steady-state analysis and optimized for the criteria outlined in Section~\ref{ssec: trim_optimization}. Shown are the trim-actuation for wing-tilt $\zeta_w^t$ (left), main-propeller throttle $\delta_{pl,r}^t$ (middle) and the trim pitch $\theta^t$ (right) as function of airspeed $v_a$ and flight-path angle $\gamma$.}
	\label{fig: TM_smooth}
	\vspace{-0.5cm}
\end{figure}

Fig.~\ref{fig: TM_smooth} shows the trim-maps for the wing-tilt angle $\zeta_w^t$, the main-throttle setting $\delta_{pl}^t=\delta_{pr}^t=:\delta_{pl,r}^t$ and the aircraft pitch $\theta^t$ obtained by the above described optimization. The trim-throttle map ($\delta_{pl,r}^t$) clearly shows that the steady flight-envelope in climb ($\gamma>0$) is limited by main-throttle saturation. Also note how the maximum possible airspeed $v_a$ reduces with increasing flight-path angles. Around hover, the trim-pitch $\theta^t$ is close to zero, whereas, in cruise-flight, pitch is aligned with the flight-path angle. This solution is close to the desired trim-pitch $\theta^*$ imposed in the optimization (Section~\ref{ssec: trim_optimization}).  
Overall, the basic operation of the UAV is well illustrated: for a forward-transition, the wing is gradually tilted down with increasing airspeed and, as soon as wing-born lift dominates, throttle is mainly used to counteract air-drag in forward flight and, therefore, it can be reduced. 

\section{ATTITUDE CONTROL}\label{sec: attitude_control}
In order to stabilize the UAV attitude in all flight phases at a desired setpoint, we develop a model-based attitude controller. Its structure is shown in Fig.~\ref{fig: att_ctrl_strct}. The setpoint consists of roll- ($\phi_{des}$), pitch- ($\theta_{des}$) and yawrate ($\dot{\psi}_{des}$) references \added{that are provided manually or by the cruise controller.} 
In the following, we present the three main parts of the control system:
\begin{figure}[b]
	\vspace{-0.5cm}
	\centering
	\includegraphics[width = 1\linewidth]{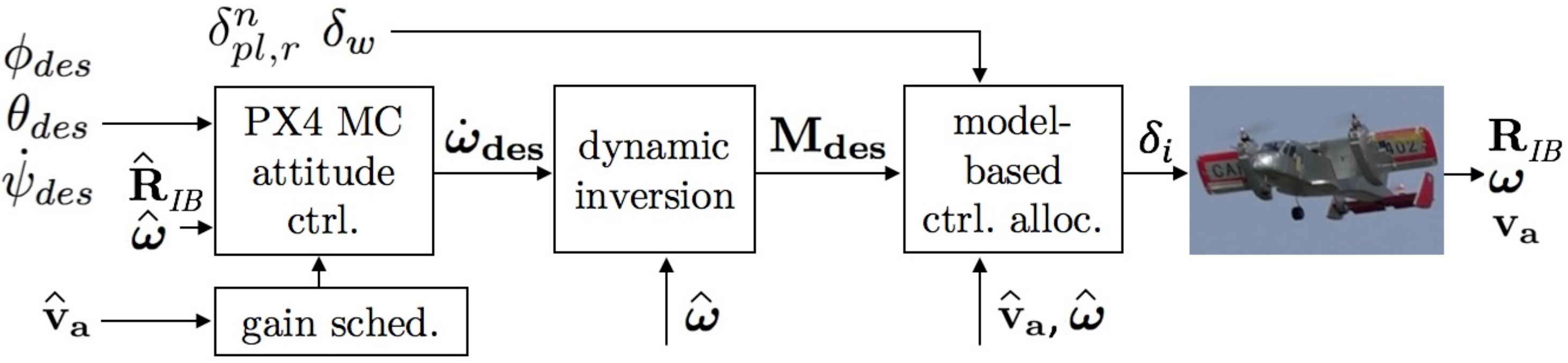}
	\caption{Architecture of the proposed attitude control system to track roll- ($\phi_{des}$), pitch- ($\theta_{des}$) and yawrate ($\dot{\psi}_{des}$) references. Wing-tilt ($\delta_w$) and nominal main-propeller throttle ($\delta_{pl,r}^n$) are commanded manually or by the \ac{CCS}. Estimated quantities are denoted by a hat.}
	%\vspace{-0.5cm}
	\label{fig: att_ctrl_strct}
\end{figure}
\vspace{-0.25cm}
\subsection{Controller}\label{ssec: attitude_controller}
The basic attitude controller is a model-free, non-linear control law based on a quaternion-formulation which maps errors in the attitude to a desired body angular acceleration $\dot{\bm{\omega}}_{\mathbf{des}}$ \added{using a cascaded P-PID-structure}. It is already implemented in the PX4 autopilot software for multirotor control and thus adopted for the present work, \cite{Meier2015}. An in-depth outline and analysis of the controller is given by the respective authors in \cite{Brescianini2013}. 
Due to the tail-propeller generating only upwards forces (negative pitch-moments) and the CG being close to the midpoint between the main-propellers, a non-symmetric pitch authority is present around hover: to improve pitch-response, a pitch-error dependent gain-scheduling is therefore included in this flight-phase\added{, i.e., selecting weaker pitch-gains when pitching down to reduce overshoot}. 
\vspace{-0.25cm}
\subsection{Dynamic Inversion}\label{ssec: dynamic_inversion}
Given the desired angular acceleration $\dot{\bm{\omega}}_{des}$ calculated in the attitude control block, the total moment $\vect{M}_{des}$ required to achieve $\dot{\bm{\omega}}_{des}$ is obtained by rearranging (inverting) the angular dynamics (\ref{eq: EoM}) of the aircraft:
\begin{equation}\label{eq: FBL}
\prescript{}{\mathcal{B}}{\mathbf{M_{des}}} =  
\prescript{}{\mathcal{B}}{\vect{I}}
\prescript{}{\mathcal{B}}{\dot{\bm{\omega}}_{\mathbf{des}}} + \prescript{}{\mathcal{B}}{\bm{\omega}}\times
\prescript{}{\mathcal{B}}{\mathbf{I}}
\prescript{}{\mathcal{B}}{\bm{\omega}}
\end{equation}
with $\bm{\omega}$ the current angular body rate. In general terminology, $\vect{M}_{des}$ is known as virtual input to the system and, by the above choice, linearizes the angular system dynamics to $\bm{\omega}=\bm{\omega}_{des}$ \cite{Khalil2002}. However, note that in practice, $\vect{M}_{des}$ can be generated only approximately due to modeling errors, state-estimation uncertainty, actuator saturation and external disturbances. The overall controller must thus be robust enough to compensate for the resulting errors in the moment. 
\vspace{-0.25cm}
\subsection{Control Allocation}\label{ssec: att_ctrl_allocation}
\begin{figure}[thpb]
	\centering
	\includegraphics[width = 0.7\linewidth]{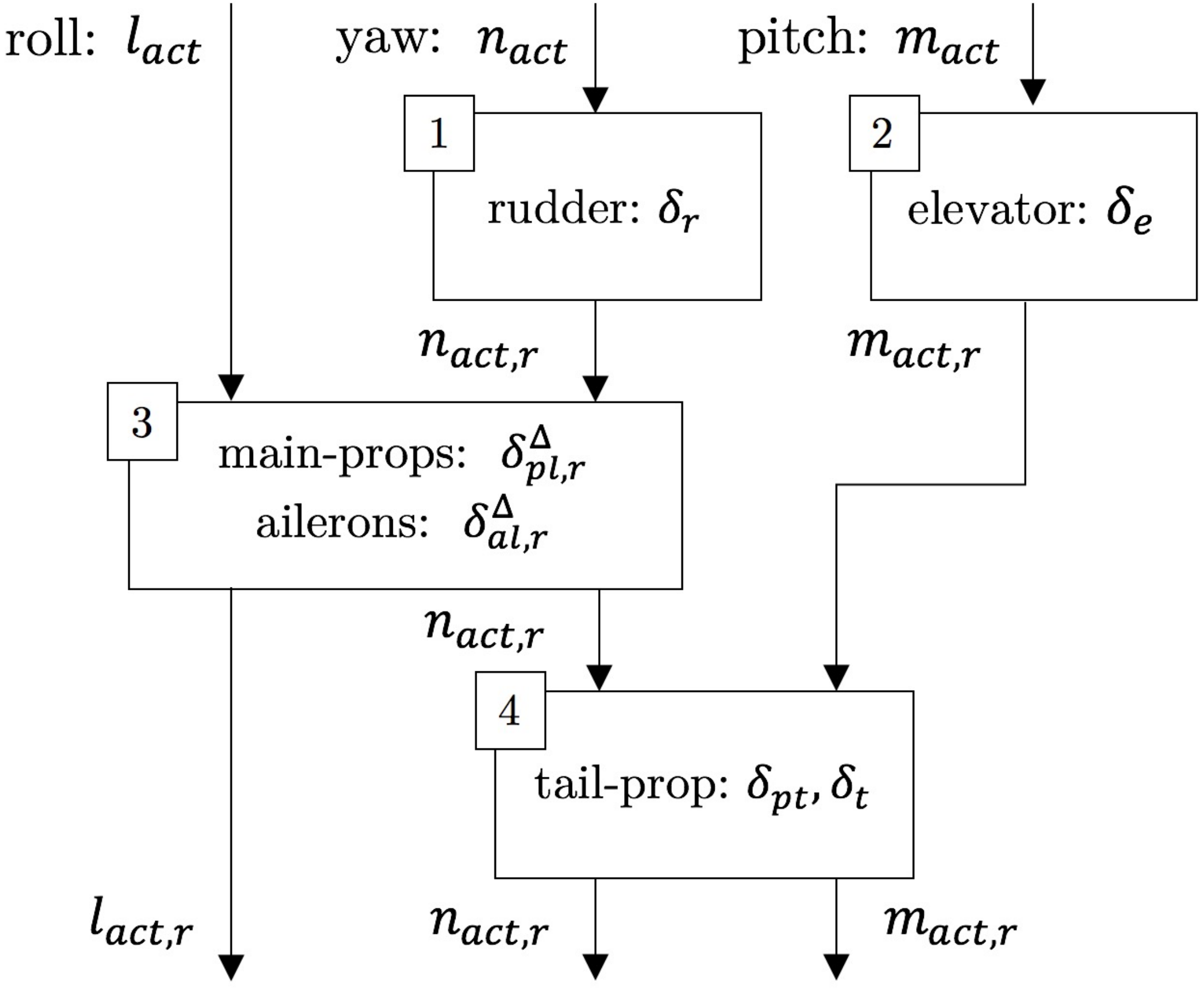}
	%\vspace{-0.25cm}
	\caption{The distribution of the total attitude control effort $\mathbf{\prescript{}{\mathcal{B}}{M_{act}}}=(l_{act},m_{act},n_{act})^T$ among redundant actuators employs the heuristic \emph{daisy chaining} approach. The subscript $()_r$ denotes residual moments left after allocation of preceding actuators in the chain, the superscript $()^\Delta$ indicates differential actuation. Due to control couplings, differential ailerons/throttle (block 3) and tail-throttle/-tilt (block 4) are allocated in groups.}
	\vspace{-0.5cm}
	\label{fig: chaining}
\end{figure}

In the last step, the system actuation required to generate $\mathbf{M_{des}}$ is determined. 
For this purpose, we define the aerodynamic moment $\mathbf{M_{act}}$ to be actuated as 
\begin{equation}\label{eq: m_act}
\prescript{}{\mathcal{B}}{\mathbf{M_{act}}} = 
\left[\begin{array}{c}
l_{act}\\
m_{act}\\
n_{act}\\
\end{array}\right] = \prescript{}{\mathcal{B}}{\mathbf{M_{des}}} - \prescript{}{\mathcal{B}}{\mathbf{\hat{M}}}(\prescript{}{\mathcal{B}}{\hat{\bm{\omega}}},\prescript{}{\mathcal{B}}{\hat{\bm{v}}_a}, \vect{u^n})
\end{equation}
where $\mathbf{\hat{M}}$ denotes the current estimate of the total aerodynamic moment acting on the vehicle with nominal actuation $\vect{u^n}$, defined by $\delta_{\lbrace al,ar,e,r,tt,pt\rbrace}^n=0$, $\delta_{pl,r}^n$ and $\delta_w$ as imposed by the pilot or the \ac{CCS}, Fig.~\ref{fig: att_ctrl_strct}. $\mathbf{\hat{M}}$ is based on the aerodynamic model and the estimated state of the UAV. Recalling the overactuated attitude of the vehicle, an approach to distribute the total control effort $\mathbf{M_{act}}$ among the actuators is required. Due to limited onboard computational power, a full online optimization is not feasible. Instead, we thus employ the lightweight heuristic known as \emph{daisy chaining} \cite{Johansen2013}. Given a defined order of priority among redundant actuators, this method sequentially allocates the actuators until the total control effort is achieved. Actuators ``further back in the chain'' remain in their nominal state. Fig.~\ref{fig: chaining} illustrates this procedure and shows the choice of priorities among the actuators and actuator-groups. Use of control surfaces is prioritized since they are considered more energy efficient than thrust vectoring. This is found to work well for the available actuators on the UAV: Pitching, e.g., is performed with the elevator in cruise and becomes gradually assisted by thrust vectoring at low speeds if the elevator saturates due to reduced effectiveness. Note that the group of actuators on the wing (ailerons and main-propellers) is the only source for roll-moment generation. Hence, the roll-axis is not overactuated and, accordingly, not daisy-chained.

With the desired control effort assigned to an actuator (or actuator group), the actual control inputs are obtained by solving the aerodynamic model for the control surface deflections and propeller speed increments, respectively. The corresponding equations are linear and quadratic in the desired variables. The result is constrained to satisfy actuator limits and then added to the nominal actuation $\delta_i^n$. In case of control saturation, non-zero residual control effort is passed on to the next actuator.

Ailerons and differential main-throttle (block 3) as well as tail-tilt and tail-throttle (block 4) are allocated in groups to handle the control couplings. Since each group provides two degrees of freedom for moment generation, actuator saturation needs to be addressed explicitly: a constrained quadratic optimization trades off roll- and yaw-moment generation on the wing (block 3), whereas attaining the pitch-moment on the tail is strictly prioritized over the yaw-moment \mbox{(block 4)}. 

Actuation of the wing-tilt and nominal main-propeller throttle is not part of the attitude controller. Instead it is commanded by either the pilot or by the higher-level \ac{CCS} (Section~\ref{sec: cruise_control}). Furthermore, for the sake of simplicity, the current implementation of the control system ignores actuator dynamics, i.e., we assume that propeller throttling and control-surface deflections are immediate. 
\vspace{-0.25cm}
\section{CRUISE CONTROL}\label{sec: cruise_control}
\begin{figure}[tb]
	\centering
	\includegraphics[width = 1\linewidth]{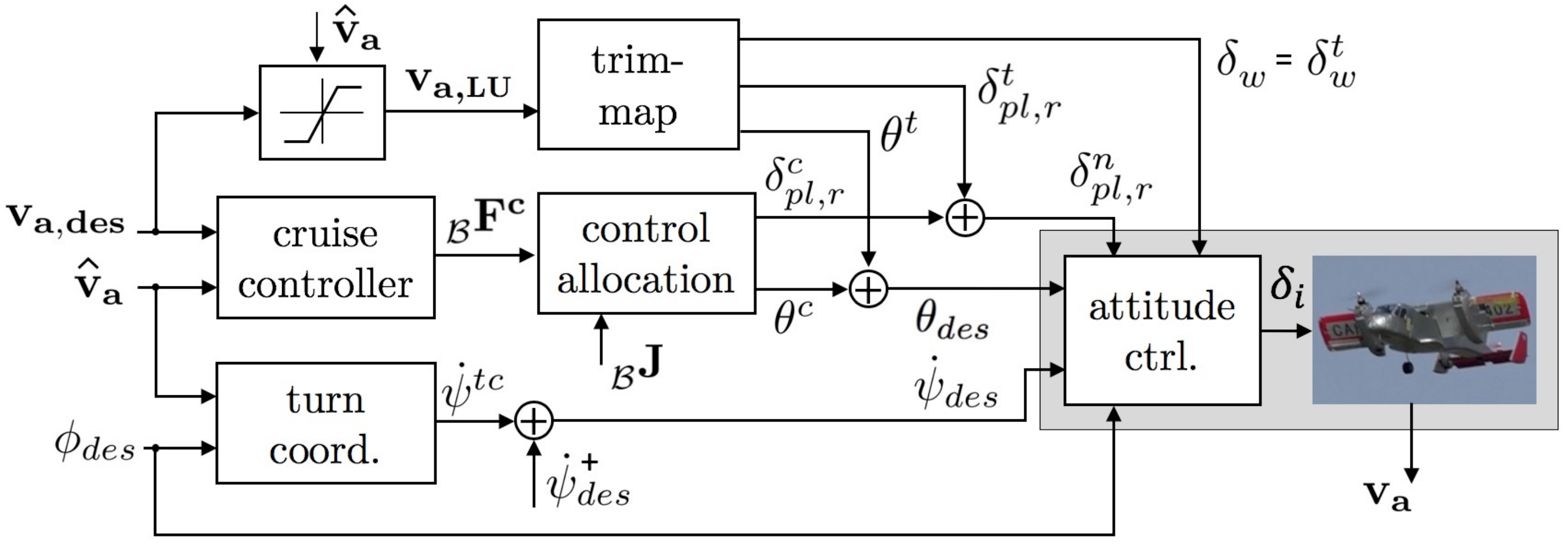}
	\caption{The architecture of the proposed cruise control system consists of two main components: 1) a feed-forward path to set the (approximate) trims for wing-tilt $\delta_w^t$, throttle $\delta_{pl,r}^t$, and pitch-angle $\theta^t$ and 2) a feedback loop to stabilize the UAV at the desired velocity $\vect{v_{a,des}}$ by commanding throttle- $\delta_{pl,r}^c$ and pitch corrections $\theta^c$. Additionally, a turn-coordination is included in cruise which mixes desired roll angle $\phi_{des}$ and yaw-rate $\dot{\psi}^{tc}$ to maintain zero lateral acceleration. Estimates are denoted by a hat.}
	\label{fig: cruise_ctrl_strct}
	\vspace{-0.5cm}
\end{figure}

With the cruise control system, operation of the UAV is further simplified by allowing the pilot to command a desired horizontal- and vertical airspeed, thus automating wing-tilt-, throttle- and pitch-angle selection. Fig.~\ref{fig: cruise_ctrl_strct} outlines the basic architecture of the cruise-control system and its interface to the attitude controller.
\vspace{-0.25cm}
\subsection{Trim-Map Feed Forward Terms}\label{ssec: trim_map_ff}
The strongly and non-linearly varying trims for throttle, wing-tilt angle and pitch angle are obtained from the trim-maps (Section~\ref{ssec: trim_analysis_results})  
and then fed forward to the attitude controller. Since the trim-maps describe the steady-state actuation, they are primarily suited for feed-forward when the airspeed setpoint is constant. 
If, however, a varying setpoint is to be tracked, the look-up-velocity ($\vect{v_{a,LU}}$, Fig.~\ref{fig: cruise_ctrl_strct}) needs to be set such that the trims i) remain close enough to the steady-state solution to retain performance of the feedback control-loop but ii) still allow to perform an unsteady maneuver. 
The former is required since the feedback control-law (Section~\ref{ssec: cc_fb_law}) is based on a local linearization of the system. Its performance thus degrades with `off-trim' feed-forward actuation. At the same time, an `off-trim' feed forward of the wing-tilt is required\footnote{Currently, we control the wing-tilt purely by feed-forward and, since it constitutes a key actuator for horizontal acceleration in early transition, trim-map look up needs to yield an unsteady actuation for acceleration.} to accelerate or decelerate during transitions. 
Given a desired airspeed vector $\vect{v_{a,des}} = (v_{a,x}, v_{a,z})^T$, the look-up in the trim-map ($\vect{v_{a,LU}}$) is thus constrained to the proximity ($v_{a,i}^{\pm}>0$) of the actual airspeed vector ($\vect{v_a}$). \deleted{We therefore require the horizontal ($i=x$) and vertical ($i=z$) components to satisfy:}\added{For the horizontal ($i=x$) and vertical ($i=z$) components we require:}
\begin{equation}
v_{a,i}-v_{a,i}^- < v_{a,i,LU} < v_{a,i}+v_{a,i}^+
\end{equation}
This method leads to a trade-off when selecting the bounds $v_{a,i}^+$,$v_{a,i}^-$: If they are set too tight, the aircraft might fail to accelerate/decelerate or does so only very slowly. On the other hand, a far `off-trim' situation might arise with the mentioned performance degradation of the stabilizing feedback controller. Acceptable values for those bounds are found in flight-testing by gradual relaxation until transitions become possible or feedback control performance degrades---the former is found to occur first if the trim-maps are accurate enough.
\vspace{-0.25cm}
\subsection{Controller}\label{ssec: cc_fb_law}
In order to correct for modeling errors corrupting the trim-maps, to attenuate disturbances, and to increase tracking performance of the desired airspeed vector $\vect{v_{a,des}}$ by providing the required `maneuvering' forces, an additional, stabilizing feedback control law is inevitable. We use a PID structure to map velocity errors to desired accelerations and---based on the mass of the \ac{UAV}---to corrective forces, respectively. Allocation of the corrective throttle $\delta_{pl,r}^c$ and -pitch $\theta^c$ follows a regularized, weighted least-squares approach based on local control derivatives $\vect{J}$: 
\begin{equation}\label{eq: CC allocation}
\begin{array}{c}
\underset{\vect{u^c}}{\text{min}}\left(
\vect{J}\vect{u^c}-\vect{F^c}
\right)^T
\vect{W}\left(
\vect{J}\vect{u^c}-\vect{F^c}
\right) + 
(\vect{u^c})^T\vect{K}
\vect{u^c}\\[3pt]

\vect{u^{c}} = 
\left(
\begin{array}{c}
\theta^{c}\\
\delta_{pl,r}^{c}
\end{array}
\right),\quad

\vect{J} = 

\left(
\begin{array}{cc}
\frac{\partial f_x}{\partial\theta}&\frac{\partial f_x}{\partial\delta_{pl,r}}\\[3pt]
\frac{\partial f_z}{\partial\theta}&\frac{\partial f_z}{\partial\delta_{pl,r}}
\end{array}
\right)\\\\

\end{array}
\vspace{-0.25cm}
\end{equation}
with $\vect{F^c}=(f_x,f_z)^T$ the desired corrective force obtained from the controller and $\vect{W}$, $\vect{K}$ symmetric, positive definite weighting and regularization matrices, respectively. The control derivatives are numerically approximated using finite differences and the aerodynamic model of the UAV. 
Contributions of stalled airfoil segments to $\partial f_i/\partial\theta$ are ignored due to modeling uncertainties\footnote{Close beyond stall-angle, lift- and drag-curves typically exhibit a hysteresis in reality. Our model ignores this fact which, in turn, is found to cause pitch-angle instabilities when employed for control-derivative calculation of airfoils close to stall.}. 
The wing-tilt $\delta_w$ is not part of this control law since its dynamics are very slow compared to the throttle and pitch-response, it thus remains being fed-forward only ($\sim\SI{5}{s}$ and $\sim\SI{10}{s}$ for fully tilting up and down, respectively).

The weighted least-squares approach allows to trade off the realization of horizontal ($f_x$) and vertical ($f_z$) corrective forces using the weighting matrix $\vect{W}$. Strong coupling effects between the corresponding axes are present in transition where, e.g., propeller thrust contributes to both $f_x$ and $f_z$. If an actuator saturates or constraints are set on maximum allowed corrective pitch $\theta^t$, simultaneous control of both axes entails degraded axis-wise performance in comparison to single-axis control. For overall safety, we thus prioritize vertical- over horizontal velocity control and ignore $f_x$ during transitions by appropriate scheduling of $\vect{W}$: 
\added{The $f_z$-weight, $w_{zz}$, remains constant and $w_{xx}$ is linearly ramped up from $w_{xx}<<w_{zz}$ to $w_{xx}~\sim w_{zz}$ as airspeed increases from \SI{12}{m/s} to \SI{15}{m/s} 
($w_{xz}=w_{zx}=0$).}

\section{EXPERIMENTAL VERIFICATION}\label{sec: experimental_verification}

The presented control-system was extensively tested both in simulation and on the real platform. We performed initial tuning of the attitude-controller gains in hover configuration and with the UAV suspended by a tether for experimental safety. 
Subsequently, we assessed the stabilization and tracking performance in outdoor experiments for all flight phases, including partial- and full transitions of various durations. Finally, the \ac{CCS} was included and investigated on vertical airspeed control capabilities.

\subsection{Attitude Control}
\begin{figure}[b]
	\centering
	\includegraphics[width = 1.0\linewidth]{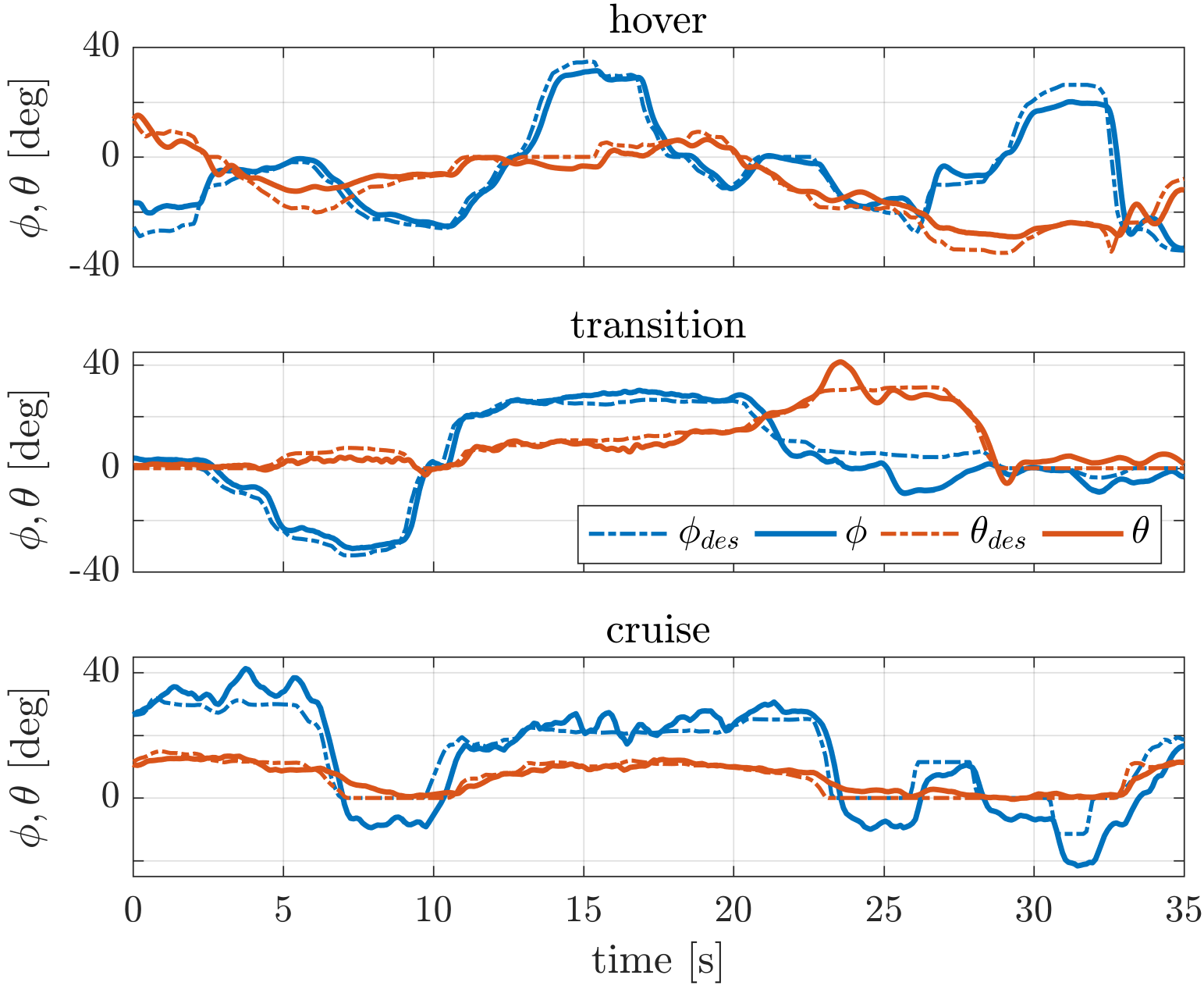}
	\vspace{-0.5cm}
	\caption{Attitude control and tracking of roll ($\phi_{des}$) and pitch angle ($\theta_{des}$) reference. Top: during hover, Middle: in transition with $\zeta_w\approx\ang{45}$ and airspeed $\sim \SI{8}{m/s}$, integrators are included on the level of roll- and pitch axis rate control. Bottom: during cruise and with airspeed $\sim \SI{20}{m/s}$ (wind amounted to $\sim\SI{8}{m/s}$, presumably turbulent), steady roll-offset is persistent as no integrators were used in this flight.}
	\vspace{-0.25cm}
	\label{fig: att_ctrl_man}
\end{figure}

Fig.~\ref{fig: att_ctrl_man} shows roll- and pitch angle tracking in hover, transition and cruise flight.  
Overall, acceptable performance is observed in most phases. However, pitch-angle tracking was found to exhibit degradation at wing-tilt angles $\zeta_w\sim\ang{20}$. There, the free-stream immersed parts of the wing are just stalled and pitching is about to require support by the tail-propeller. Strong non-linearities due to complex aerodynamic effects and modeling errors can explain the observed degradation. 

Attitude control during transitions did consider in particular stabilization of the pitch-angle: a fast back-transition, i.e., going from cruise to hover, is depicted in Fig.~\ref{fig: cc_level_BT}. As seen, \added{pitch is disturbed in this regime but} the pitch error can be kept below $\ang{5}$ most of the time. Note that roll-axis control is generally less demanding due to the symmetry of the system and thus performs better than pitch. 

\subsection{Cruise Control}
\begin{figure}[tb]
	\centering
	\includegraphics[width = 1.0\linewidth]{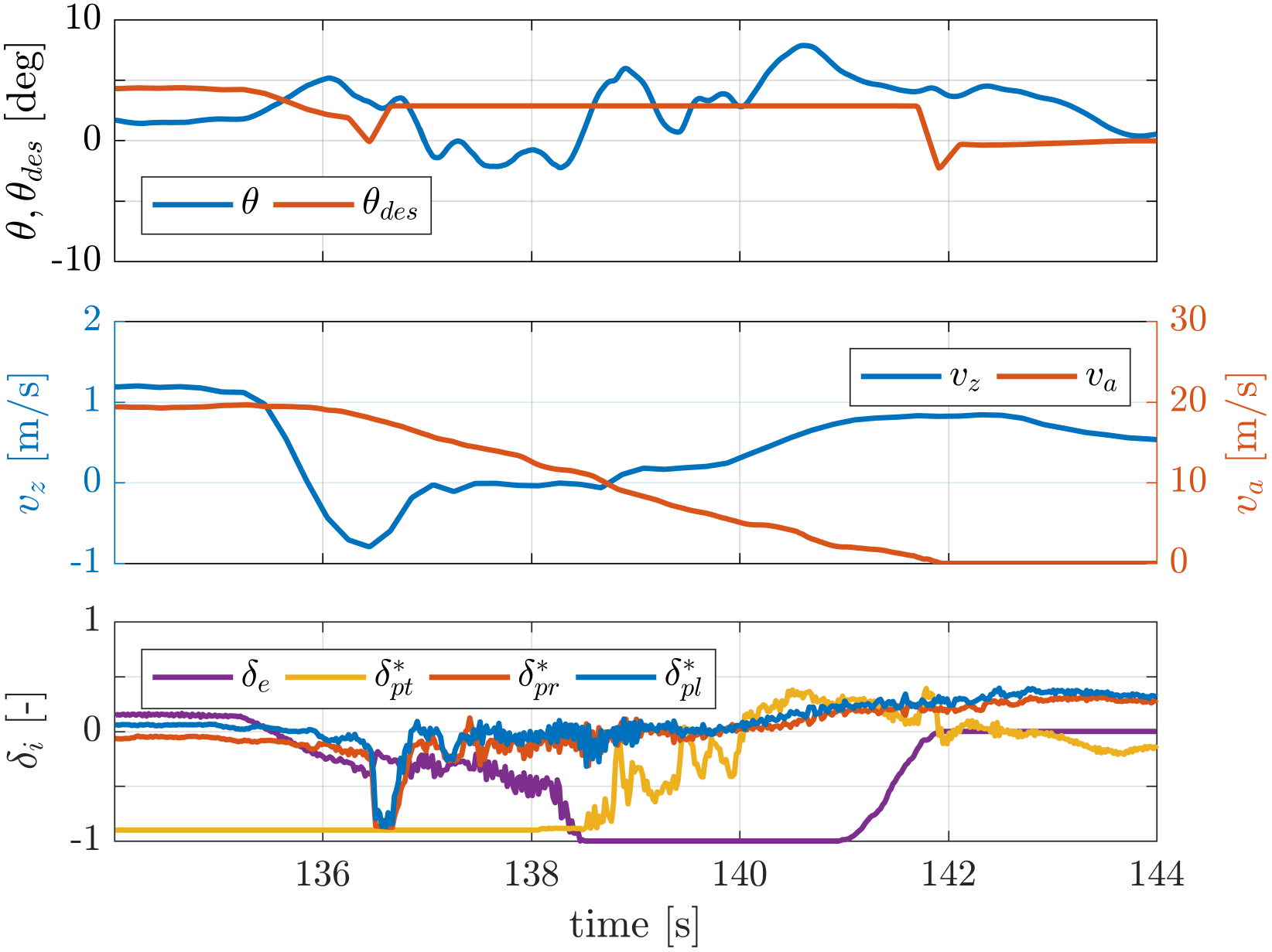}
	%\vspace{-0.5cm}
	\caption{Back-transition with cruise controller commanded to keep zero vertical velocity $v_z$. Total altitude difference for full transition $<\SI{2}{m}$. Bottom plot shows the main actuator inputs (normalized) used to control the transition, notice the combined actuation of the tail-prop and the elevator to generate a negative pitch-moment (cf. daisy chaining, Section.~\ref{ssec: att_ctrl_allocation}). The main-throttle is minimal in the first phase of the back-transition to keep the climb-rate low. The range of normalized throttle-inputs is stretched from $[0,1]$ to $[-1,1]$ for ease of display ($\delta_i^*$).
	}
	%\vspace{-0.25cm}
	\label{fig: cc_level_BT}
\end{figure}

\begin{figure}[tb]
	\centering
	\includegraphics[width = 1.0\linewidth]{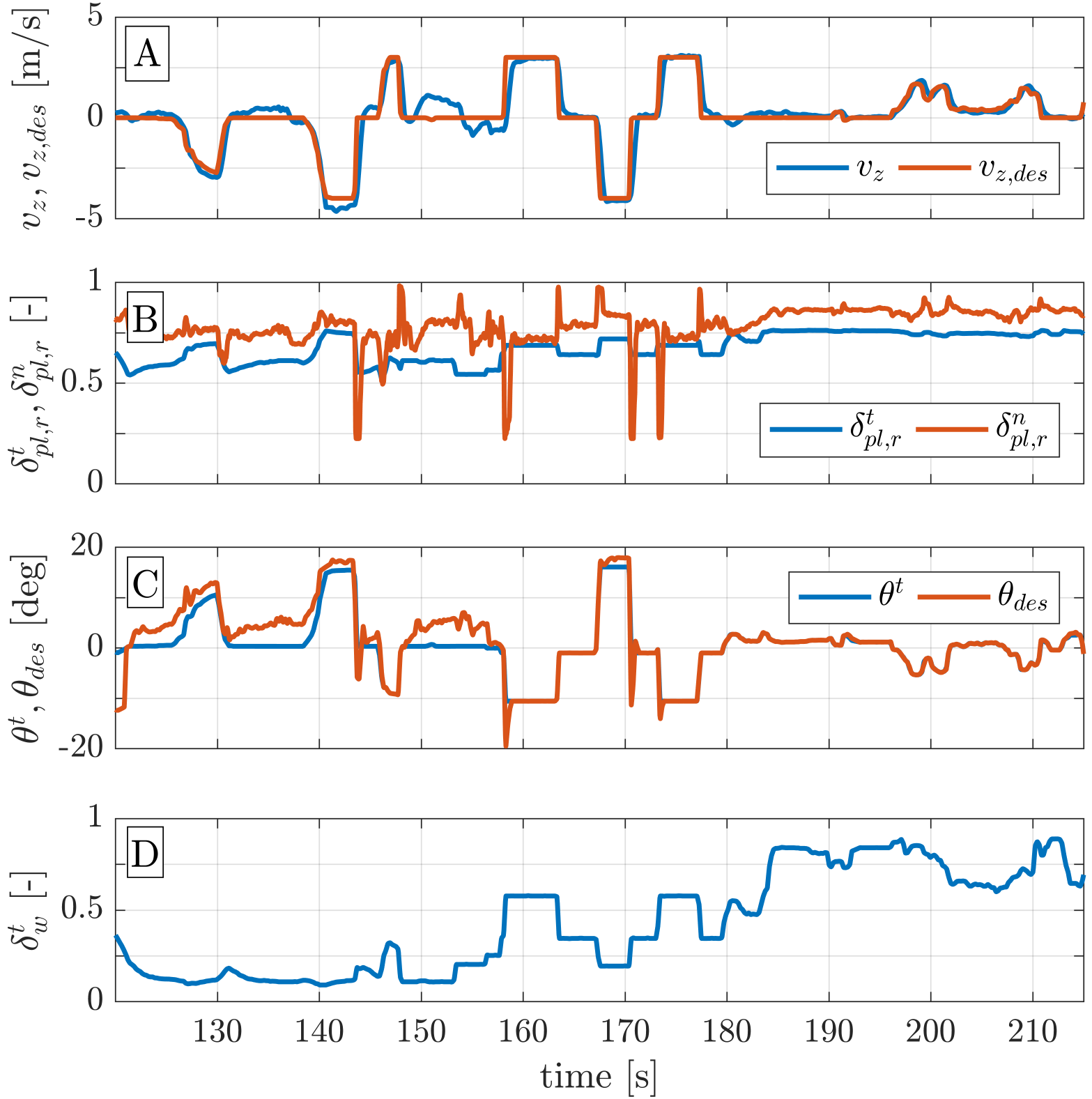}
	%\vspace{-0.5cm}
	\caption{Vertical velocity control subject to multiple step inputs: A) Reference and actual $v_z$ showing good tracking, error at $t=\SI{150}{s}$ during a turn with high roll-angle $\phi\approx40^\circ$ which is currently not compensated for. B) Throttle setting obtained from trim-map $\delta_{pl,r}^t$ and commanded nominal throttle $\delta_{pl,r}^n$, the corrective throttle is given as $\delta_{pl,r}^c = \delta_{pl,r}^n-\delta_{pl,r}^t$, C) as in B but for the pitch-angle, D) Feed forward trim wing-tilt $\delta_w^t$ obtained from the trim-map.}
	\vspace{-0.25cm}
	\label{fig: cc_vz_ctrl}
\end{figure}

Fig.~\ref{fig: cc_level_BT} demonstrates vertical velocity stabilization during a fast back-transition which is generally characterized by highly non-linear and fast varying dynamics of the UAV: a strong increase in lift and positive pitch moment is followed by a sudden decrease of the same when reaching stall. On-time throttling is therefore key for vertical velocity control. The presented controller is able to maintain altitude within a $\SI{2}{m}$ band while fully decelerating from \mbox{$\SI{20}{m/s}$} in \mbox{$\sim\SI{6}{s}$}.
Vertical velocity tracking is demonstrated in Fig.~\ref{fig: cc_vz_ctrl}A for maximum $v_z$ step-inputs and gradually varying horizontal velocity ($v_x\sim\SI{15}{m/s}-\SI{3}{m/s}$), showing accurate tracking and short response times. Comparing feed-forward and corrective values for throttle and pitch reveals the merits of both the trim-map and the feedback controller: For, e.g., $\delta_{pl,r}^n$, the trim-map contributes up to $80\%$ of the control signal in steady-state phases while the feedback terms dominate during unsteady phases to achieve the fast responses upon setpoint changes.

\section{FUTURE WORK}\label{sec: conclusion_future_work}

Further steps in the development of the control system consider a proper system identification \added{and/or wind-tunnel testing} to obtain those parameters of the aerodynamic model which are, for now, only based on typical values from literature. Given the strong dependence of the proposed attitude- and cruise controller on the aerodynamic model, we expect controller performance to benefit from more accurate parameter values. Furthermore, modifications of the tilting mechanism of the wing towards a faster and more reliable actuation would allow to include the wing-tilt in the cruise-control feedback-loop. The added control authority would simplify simultaneous horizontal- and vertical cruise control in the transition phase. 


%%%%%%%%%%%%%%%%%%%%%%%%%%%%%%%%%%%%%%%%%%%%%%%%%%%%%%%%%%%%%%%%%%%%%%%%%%%%%%%%

%%%%%%%%%%%%%%%%%%%%%%%%%%%%%%%%%%%%%%%%%%%%%%%%%%%%%%%%%%%%%%%%%%%%%%%%%%%%%%%%

%%%%%%%%%%%%%%%%%%%%%%%%%%%%%%%%%%%%%%%%%%%%%%%%%%%%%%%%%%%%%%%%%%%%%%%%%%%%%%%%

%\section*{APPENDIX}

\deleted{
\section*{ACKNOWLEDGMENT}
The authors would like to thank Dufour Aerospace (https://dufour.aero) for initiating and supporting this project.}
%%%%%%%%%%%%%%%%%%%%%%%%%%%%%%%%%%%%%%%%%%%%%%%%%%%%%%%%%%%%%%%%%%%%%%%%%%%%%%%%

\balance
\bibliographystyle{IEEEtran}
\bibliography{IEEEabrv,root}

\end{document}